\documentstyle[12pt,twoside,fleqn,espcrc1]{article}

\newcommand{\AmS}{{\protect\the\textfont2
  A\kern-.1667em\lower.5ex\hbox{M}\kern-.125emS}}
\newcommand{\gapproxeq}{\lower.7ex\hbox{$\;\stackrel{\textstyle>}{\sim}\;$}}
\newcommand{\arrowslash}{\lower.1ex\hbox{$\;\stackrel{{\tiny not}}{\rightarrow}\;$}}
\newcommand{\bnabla} {{\mbox{\boldmath $\nabla$}}}
\newcommand{\bsigma} {{\mbox{\boldmath $\sigma$}}}
\newcommand{\bx} {{\bf x}}
\newcommand{\ba} {{\bf A}}
\input{psfig.sty}

\title{ \vspace{-2cm}\begin{flushright} 
\small{hep-ph/9909201}
\\ \small{LA-UR-99-4707}
 \end{flushright} \vspace{2cm}
Gluonic Excitations' Millennial Finale}

\author{Philip R. Page\footnote{Invited talk at XV$^{th}$ Particles and Nuclei International Conference (PANIC 99), Uppsala, Sweden, 10--16 June 1999.}\address{Theoretical Division, MS-B283, Los Alamos
National Laboratory, \\ P.O. Box 1663, Los Alamos, NM 87545, USA}}
       
\begin{document}

\maketitle

\begin{abstract}
We provide an overview of theoretical developments on 
hybrid mesons and glueballs in the last year at this turn of the
millenium conference. Cracks in potential models of conventional
mesons are developing. Hybrid meson adiabatic surfaces have been
calculated and interpreted, experimental $J^{PC}$ exotics
have hybrid meson, four--quark state or non--resonant interpretations,
and the strong decay mechanism of hybrids has been studied. All 
theoretical progress on hybrid mesons in the last year is mentioned. 
Overall features of glueballs 
are visited: decays and the 
successes of the large $N_c$ limit. Two promising experimental areas 
are mentioned: charmonium hybrids at $B$--factories and 
 $s\bar{s}$ hybrids at Jefferson Lab.
\end{abstract}

\section{Cracks in Potential Models: Radial Excitations}

There are three mass regions in which in which there are recent 
experimental indications for states at masses inconsistent with 
potential model predictions:

\begin{itemize}

\item Crystal Barrel, VES, ARGUS and L3 \cite{burakovskytensor}
have evidence for $a_2(1660)$ approximately 170 MeV lower than 
potential model expectations \cite{isgur85}.

\item DELPHI, but {\it not} CLEO and OPAL,  observes
a new resonance $D(2637)$ with a total width of less than 15 MeV \cite{pene}.
The observed mass uniquely identifies the state as a radially excited
$D^{\ast}$ if mass predictions from potential models are used 
\cite{isgur85,ebert}. However, this interpretation has been found to be
inconsistent with the narrow width \cite{pene}. Assuming that 
the DELPHI result is confirmed, this may signal a crisis for potential models.

\item  Numerous light quark meson states $\gapproxeq 2$ GeV have been 
reported recently by E818, VES and A.V. Anisovich, D.V. Bugg, 
A.V. Sarantsev and B.-S. Zou \cite{burakovskytensor}, typically at 
appreciably lower masses than found by potential models \cite{isgur85}.

\end{itemize}

At least some of these discrepancies with potential models should be 
related to the appearance of new degrees of freedom in the QCD spectrum:
hybrid mesons and glueballs.

\section{Hybrid Mesons}

\subsection{Hybrid Adiabatic Surfaces}

A careful lattice QCD calculation has been performed to map out the
energy of a $q\bar{q}$ pair, with the quarks stationary, as a 
function of the distance $r$ between them, called the {\it adiabatic 
surface} (see Figure 1) \cite{mornhybrid}. The lowest surface corresponds to 
conventional mesons, but the
excited surfaces to various hybrid mesons. The interpretations of this
result are consistent between various authors. 
At small $r$, the adiabatic bag model where quarks are
stationary \cite{mornbag}, or a glue--lump model where quarks are fixed at 
the centre of a bag \cite{karl}, gives a reasonable description of the
lattice data. At large $r$, a constituent gluon model (related
to the bag model) is not applicable \cite{swanson99}, but a
Nambu--Goto string picture instead \cite{olsson}. It has been argued that
the string behaviour only sets on at $r$ beyond that relevant for most of
hadron spectroscopy \cite{mornbag}. 

\begin{figure}
\center
\vspace{-2cm}
\psfig{figure=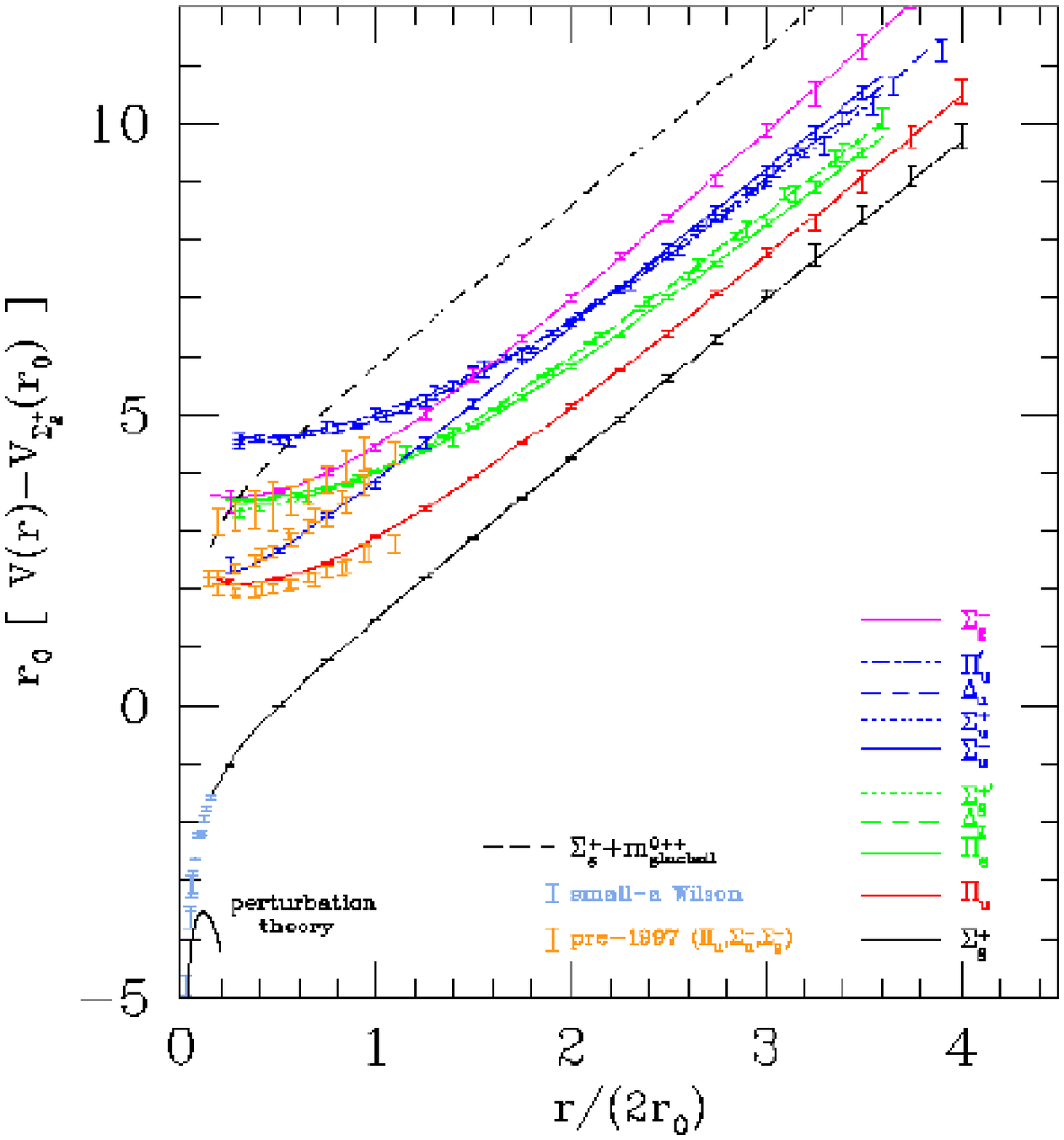,height=3.4in}
\vspace{-8.8cm}
\hspace{8.2cm}\psfig{figure=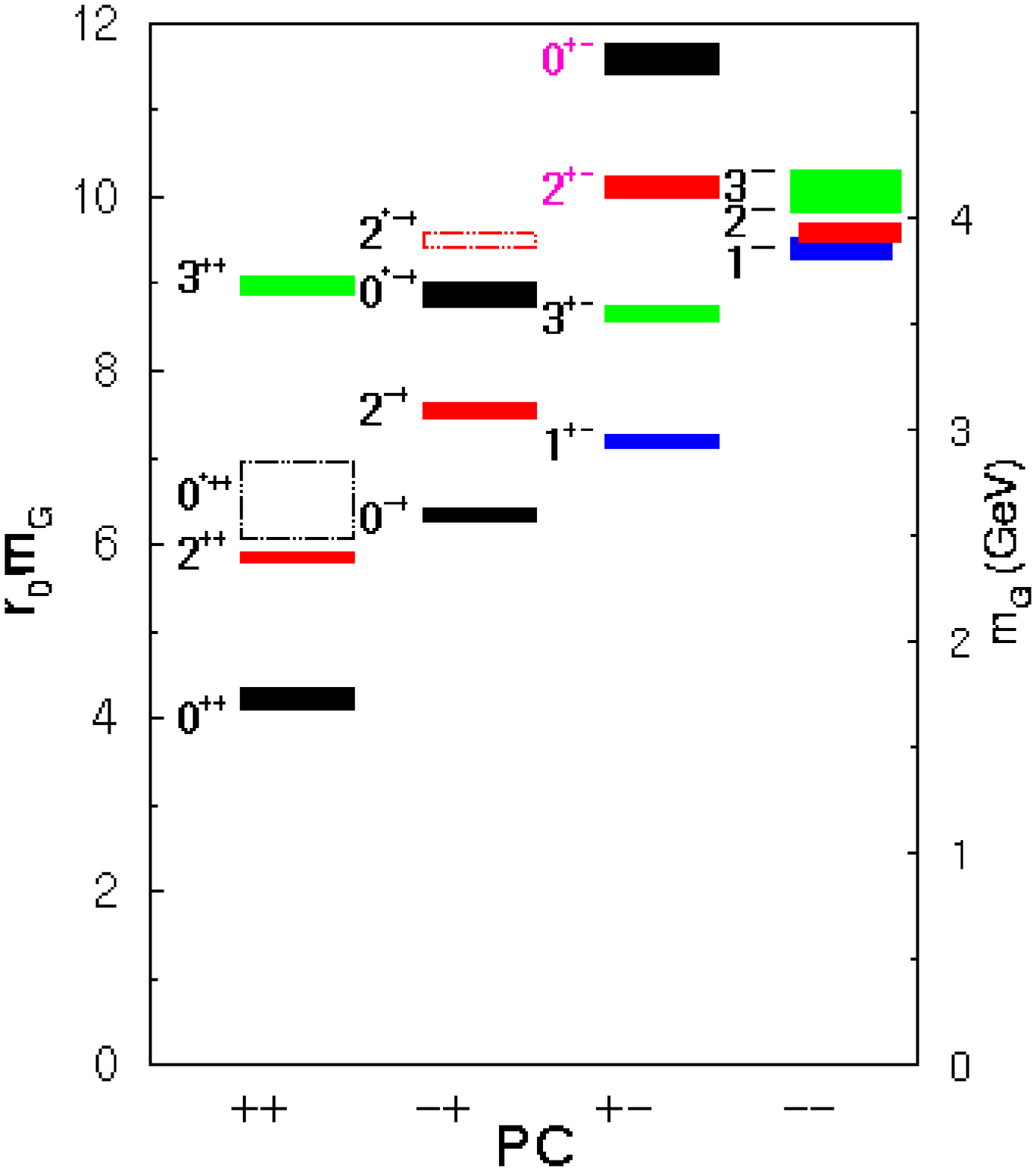,height=3.4in}
\vspace{-1.5cm}
\caption{Lattice QCD adiabatic surfaces (left) and glueball spectrum (right).}
\end{figure}

\subsection{$J^{PC}$ Exotics and their Interpretations}

Experimental isovector $J^{PC}=1^{-+}$ 
exotic mesons have been reported at $\sim 1.4$ GeV in $\eta\pi$
by E852 and Crystal Barrel, and
at $\sim 1.6$ GeV in $\rho\pi,\; \eta^{'}\pi$ and $b_1\pi$ by
E852 and VES \cite{seth}. An experimental search is currently
underway at Jefferson Lab:  for the $\eta\pi$ and $\eta^{'}\pi$
channels the spin formalism for analysis is identical to that of
a vector meson \cite{schilling}, which differs from the exotic 
by its C--parity.

The lowest state at $\sim 1.4$ GeV appears rather low in comparison to 
lattice QCD estimates for the lowest $1^{-+}$ state, which can be a hybrid 
or four--quark state. In the last year these are
$2.0(1),\; 2.1(1)$ GeV \cite{milc} and $1.9(2)$ GeV \cite{lacock}.
The latter result incorporates some effects of unquenching, increasing
the error bar. 
One has to remember that the lattice QCD predictions correspond to a
strange quark, so that the light quark states should have even lower mass,
taking experimental conventional meson masses as a guide.
For a more comprehensive discussion of lattice QCD results, 
the reader is referred to ref. \cite{mcneile99}. 

We shall now discuss three interpretations of the exotic states.

I. The $1.4$ GeV state is a ground state hybrid, and the
 $1.6$ GeV state an excited hybrid. This hypothesis
is catastrophic \cite{donnachie}. 
The low mass of the $1.4$ GeV state is a problem. 
The fact that the low mass state has only been
seen in $\eta\pi$ is qualitatively in contradiction with 
symmetrization selection rules
 that suppress the decay of $1^{-+}$ hybrids to $\eta\pi$
\cite{sel}. Within the flux--tube model, the mass difference between the
states is too small, the total width of the $1.4$ GeV state too small and
the ratio of widths of the two states opposite to experiment \cite{donnachie}.

II. The 1.4 GeV state is a four--quark state ({\it or} molecule) and the
1.6 GeV state a ground state hybrid.
First the 1.4 GeV state. The low mass is still a problem. 
It's observed decay to $\eta\pi$ is consistent with 
the symmetrization selection rules \cite{sel}, 
and with a four--quark SU(3) selection
rule that predicts $\Gamma(\eta\pi) > \Gamma(\eta^{'}\pi)$ \cite{lipkin}.
A possible problem here is that no stable $J^{PC}=1^{-+}$ four--quark
state has been found in a quark model search \cite{semay}.
For the 1.6 GeV state, its non--detection in $\eta\pi$ is consistent
with the symmetrization selection rules \cite{sel}, and 
$\Gamma(\eta\pi) < \Gamma(\eta^{'}\pi)$ with the hybrid SU(3) selection
rule \cite{lipkin}. Moreover, the decay to $\rho\pi$ is also 
according to expectations
\cite{page97}. 

III. The 1.4 GeV ``state'' is produced by the interference of a 
non--resonant Deck background peaking near 1.4 GeV with a hybrid meson
resonance at
1.6 GeV. A K--matrix fit to the E852 event and phase data in $\eta\pi$ 
 using this hypothesis in shown in Figure 2 \cite{donnachie}. 
A key feature is that the phase motion in the E852 $\eta\pi$ data at 1.4 GeV
can arise from a $1.6$ GeV resonance which does not couple
strongly to $\eta\pi$, through coupled channel effects. The Deck process
involves $\pi p$ inelastic scattering into $\eta\pi p$ via $\rho$
exchange. Deck processes are however only operative in $\pi p$ 
experiments (E852 and VES), and not in $p\bar{p}$ annihilation 
at Crystal Barrel. Hence this explanation does not explain all the
data.

\begin{figure}
\center
\vspace{-2cm}
\psfig{figure=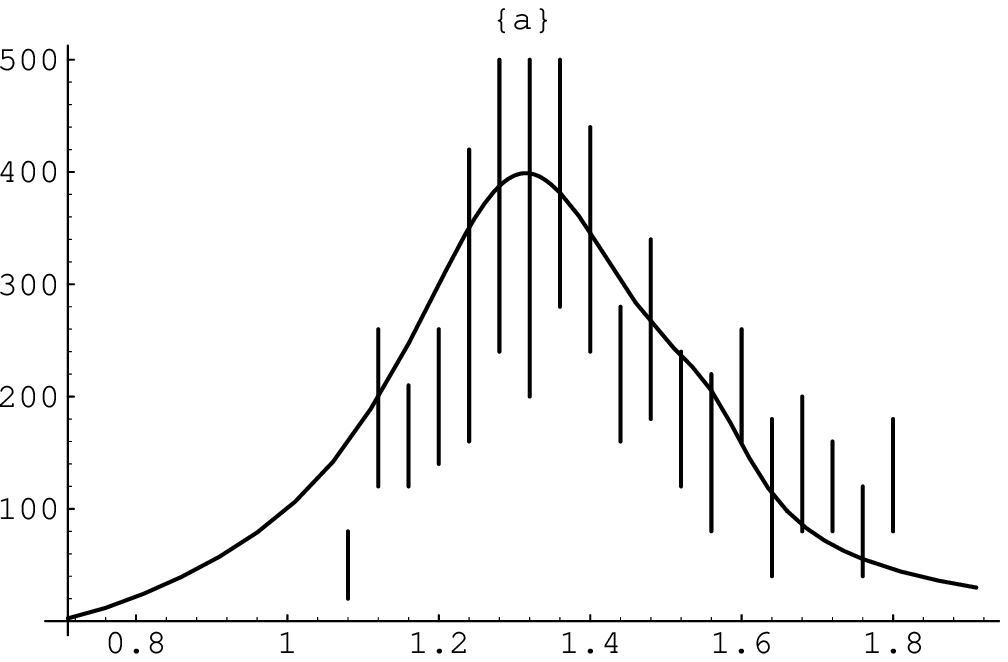,height=3.0in}
\vspace{-7.65cm}
\hspace{8cm}\psfig{figure=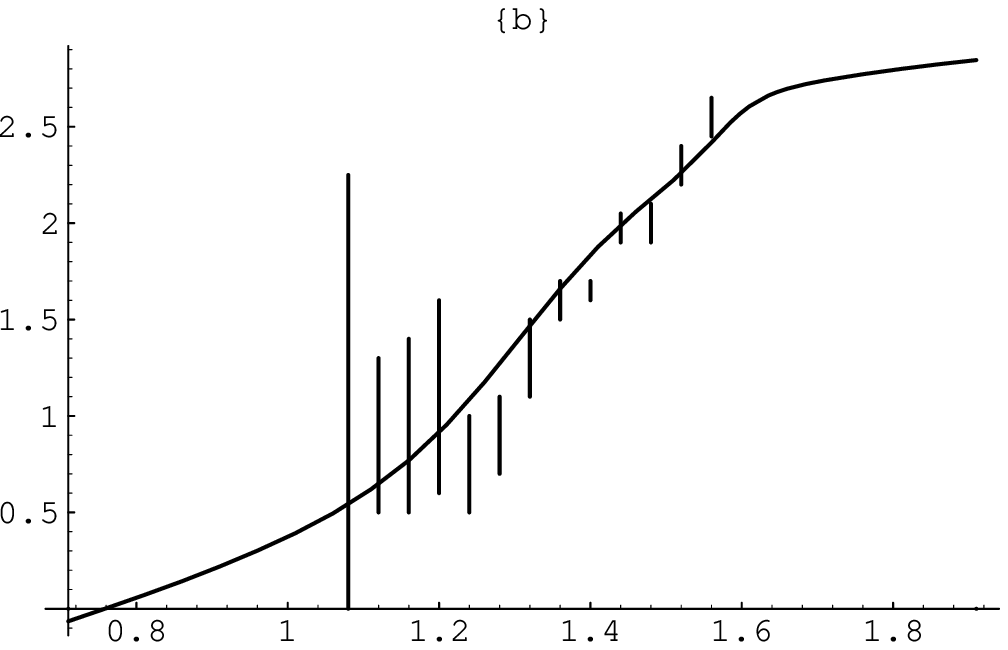,height=3.0in}
\vspace{-2cm}
\caption{ Results of the K--matrix analysis.  (a) The events 
in $\eta\pi$ as compared to experiment;
(b) The phase in $\eta\pi$ compared to experiment.
The invariant mass is plotted on the horisontal axis in GeV. }
\end{figure}

\subsection{Hybrid Strong Decay Mechanism}

The mechanism of strong decay of hybrids to two mesons via pair creation 
is of significant interest, since it may be different from that of
conventional mesons due to the structure of the glue. The first
possibility is that hybrids decay similar to the phenomenologically
successful way for mesons: 
$^3P_0$ pair creation by flux--tube breaking \cite{kokoski85}.
A second possibility, found in the heavy quark limit of QCD in Coulomb 
gauge, is $^3S_1$ pair creation by flux--tube breaking arising from the
hamiltonian \cite{swanson97}

\begin{equation}
H_{int} = \int d^3x\; \Psi^+(\bx)\; \bsigma\cdot(i\bnabla + g \ba)\; \Psi(\bx)
\end{equation}

These two pictures lead to characteristic differences between the
D-- to S--wave width ratios of $1^{-+}\rightarrow b_1\pi,\; f_1\pi$:
the ratios are much smaller in the $^3S_1$ than in the $^3P_0$ model
\cite{pss}. One may
soon be able to test these ratios with new data from E852 and VES.

\subsection{Other new results for hybrids}

The current status of light vector hybrids was discussed
\cite{kalashnikova}. Also, there were various new QCD sum rule results.
The first calculation of the width of a 1.6 GeV exotic to $f_1\pi$
yielded  
$\Gamma(1^{-+}\rightarrow\rho\pi,\; f_1\pi)=$ $40\pm 20,\; 100\pm 50$ MeV
\cite{zhu99}. Non--exotic $0^{++}$ and $0^{-+}$ light hybrid masses 
were also investigated \cite{huang99}. Also, heavy--light hybrid masses
\cite{zhu98} and decays \cite{huang98} were studied.

\section{Glueballs}

The high precision and almost complete glueball 
spectrum below $\sim 4$ GeV in Figure 1 was calculated in lattice QCD
\cite{mornglue}. In order to connect this spectrum with experiment,
mixing with and decay to conventional mesons need to be understood.

A first attempt at understanding the decay of the primitive
(bare) scalar  glueball
to pseudoscalar mesons was made some time ago in lattice QCD
\cite{sexton}. An unusual decay pattern was found, where the coupling of the
glueball to $\eta\eta$ was greater than its coupling to $K\bar{K}$,
and the latter greater than its coupling to $\pi\pi$.
Historically the expectation based on perturbative QCD was a 
``flavour democratic'' decay pattern where each of these couplings are
the same \cite{cui}. 
The first successful model to explain the lattice QCD decay
pattern postulates that the glueball does not decay directly, but only
via an intermediate scalar meson, which then subsequently decays to
the two pseudoscalars via the familiar $^3P_0$ pair creation 
mechanism \cite{burakovskyscalar}. A perfect one--parameter fit to the
three lattice points is obtained. It has  
paradoxically been found that perturbative QCD decay of glueballs
with bag model
gluon wave functions can also fit the lattice points
\cite{strohmeier}, although the indirect decay through intermediate mesons
is still present. The problem here is that perturbative QCD is not 
applicable for the scalar glueball \cite{cao}. 

The physical glueball does not necessarily 
decay flavour democratically or is narrow. This is already experimentally
evident for the $f_0(1500)$ and $f_J(1770)$. It has also recently been
found in an analysis of the tensor glueball \cite{burakovskytensor}, 
making a narrow flavour democratically decaying $f_J(2220)$ implausible.
The only hope for finding narrow glueballs appears to be when they are
unmixed with mesons. This is a possibility for glueballs beyond
3 GeV, since light mesons in this mass region may become unstable due to 
QCD pair creation, and charmonium have limited mixing
\cite{brisudova}. For example, a ``glue--rich'' quark--gluon plasma 
can produce a narrow vector glueball which decays to readily detected 
 $e^+e^-,\;\mu^+\mu^-,\;\tau^+\tau^-$. Another tantilizing 
possibility is production
of high--mass glueballs from the proposed GSI Darmstadt upgrade. 
The  caveat is that even pure glueballs may not be narrow, as lattice 
QCD finds a scalar glueball 
width of $108\pm 28$ MeV to pseudoscalars only, while
models suggest that additional modes may increase the width to at least
$250-390$ MeV \cite{burakovskyscalar}.

Interesting new results were obtained in the large number of colours 
$N_c$ limit of QCD. Teper noticed it to be an excellent guide to glueball
masses even for $N_c=3$ \cite{teper}. One can argue that the glueball mass
follows the relation

\begin{equation}
\mbox{Glueball mass }_{  N_c} = \mbox{Glueball mass }_{ { N_c}=\infty} +\frac{constant}{ { N_c}^{2}}
\end{equation}
so that corrections from the large $N_c$ limit are of 
second order \cite{teper}.
Simulations of the scalar and tensor glueball masses in SU(2), SU(3) and
SU(4) are indicated in Table 1. The agreement between $N_c = 2-4$ is
striking.

\begin{table}[tbh]
\caption{Scalar and tensor glueball masses in string tension $\sigma$ units 
 ($a\sigma=0.228\pm 0.007)$ \protect\cite{teper}.}
\label{table:1}
\newcommand{\m}{\hphantom{$-$}}
\newcommand{\cc}[1]{\multicolumn{1}{c}{#1}}
\renewcommand{\tabcolsep}{2pc} 
\renewcommand{\arraystretch}{1.2} 
\begin{tabular}{@{}llll}
\hline
 &  SU(2) & SU(3) & SU(4)  \\
\hline
$M_{0^{++}}/\sqrt{\sigma}$  & $3.62\pm 0.07$  & $3.22\pm 0.04$  & $3.28\pm 0.30$ \\
$M_{2^{++}}/\sqrt{\sigma}$  & $5.58\pm 0.03$  & $5.02\pm 0.04$  & $5.31\pm 0.48$ \\
\hline
\end{tabular}\\[2pt]
\end{table}

\section{Experiment}

\subsection{Charmonium Hybrids at $B$--factories}

The process $B\rightarrow (c\bar{c}\;\mbox{hybrid})\; X$
is an experimental frontier \cite{dunietz}. 
NRQCD estimates suggest a tiny decay branching ratio for the process
\cite{petrov}, but the $\chi_{c2}$ has already been detected at this
level \cite{dunietz}. Isgur recently argued that hybrid mesons should
have similar production to conventional mesons in factorizable processes,
which this one is not \cite{isgur99}.

One of the issues critically determing the feasibility of detecting
hybrid charmonium is the width of these states. Various arguments
suggest that they should be narrow if they are below the $D^{\ast\ast}D$
threshold, e.g. the $0^{+-}$ and $1^{-+}$ exotic hybrids should have total
widths of order $5$ and $20$ MeV respectively \cite{dunietz}.
Mass determinations for charmonium $1^{-+}$ hybrids
in the last year 
are shown in Table 2, where the first four are lattice QCD results.
The difference between the $D^{\ast\ast}D$ threshold and the  
spin--averaged ground state charmonium mass is 1220 MeV, so that current
$1^{-+}$ hybrid mass estimates straddle the $D^{\ast\ast}D$ threshold.

\begin{table}[tbh]
\caption{Difference between $1^{-+}$ hybrid 
and spin--averaged ground state quarkonium mass (MeV).}
\label{table:2}
\newcommand{\m}{\hphantom{$-$}}
\newcommand{\cc}[1]{\multicolumn{1}{c}{#1}}
\renewcommand{\tabcolsep}{2pc} 
\renewcommand{\arraystretch}{1.2} 
\begin{tabular}{@{}lll}
\hline
 $c\bar{c}$ &  $b\bar{b}$ &  \\
\hline
 $1340^{+60}_{-150}+\mbox{sys} $ &  & MILC Collab. \protect\cite{mcniele98}        \\
 $1323\pm 13 $  &  $1542\pm   8 $    & CP--PACS Collab.   \protect\cite{manke99}             \\
 $\sim 1190$   &   $1490\pm 20\pm 50$     & \protect\cite{juge}               \\
                  &  $1680\pm 100$& UKQCD Collab. \protect\cite{manke98}              \\
 $\sim 950$        &   $\sim 1200 $     & Constituent gluon model \protect\cite{gerasimov}            \\
                      &  $1300\pm 100$   & QCD sum rules \protect\cite{zhu98}   \\       
\hline
\end{tabular}\\[2pt]
\end{table}

It evident from Table 2 that the difference in mass between the $b\bar{b}$ 
hybrid and its spin--averaged ground state is {\it larger} than that for
$c\bar{c}$. This is {\it opposite} to the behaviour of conventional
radially and orbitally excited quarkonia.

\subsection{$s\bar{s}$ hybrids at Jefferson Lab}

We have performed a new study in the flux--tube model \cite{kokoski85,pss}
of the decay of the $s\bar{s}$ $1^{-+}$ exotic hybrid meson, which 
differs from earlier work in that the $s\bar{s}$ hybrid is at different 
masses. The first mass chosen is 250 MeV above the experimental $1.6$ GeV 
exotic state, i.e. $1840$ MeV, making it the $s\bar{s}$ partner of the
$1.6$ GeV state; and the second is the mass of the
exotic meson candidate $X(1920)$ \cite{pdg98}. 
The calculation is displayed in the
narrow resonance approximation in Table 3.

\begin{table}[tbh]
\caption{Partial decay widths of an $s\bar{s}$ $1^{-+}$ hybrid (MeV)
at $1840$ and $1920$ MeV in the $^3P_0$ decay model (IKP)
\protect\cite{kokoski85} and the $^3S_1$ model (PSS) \protect\cite{pss}. The
$\eta^{'}\eta$ partial width vanishes and $K^{\ast}(1410)K$ is tiny.}
\label{table:3}
\newcommand{\m}{\hphantom{$-$}}
\newcommand{\cc}[1]{\multicolumn{1}{c}{#1}}
\renewcommand{\tabcolsep}{2pc} 
\renewcommand{\arraystretch}{1.2} 
\begin{tabular}{@{}lrrrr}
\hline
Decay Mode & \multicolumn{2}{c}{\underline{$1840$ MeV}} & \multicolumn{2}{c}{\underline{$1920$ MeV}}\\ 
 & PSS & IKP & PSS & IKP \\
\hline
$K^{\ast}K$ & 9 & 6 & 10 & 7 \\
$K_1(1270)K$ & 7 & 17 & 10 & 24 \\
$K_1(1400)K$ &   &    & 31 & 73 \\
\hline
Total & 16 & 22 & 52 & 104 \\
\hline
\end{tabular}\\[2pt]
\end{table}

The salient feature is that $K_1(1400)K$ is the would--be dominant mode,
but that the $s\bar{s}$ hybrid mass straddles its threshold. 
Removing the narrow resonance approximation is 
expected to increase the width to $K_1(1400)K$.
We obtain a surprisingly narrow state, whose simplest search channel
is $K^{\ast}K$. A natural place to search for $s\bar{s}$ hybrids
is at Jefferson Lab, where the photon has a sizable coupling to 
$s\bar{s}$. However, production may be complicated, as diffractive
exchange is forbidden by C--parity conservation, and meson exchange
involves OZI forbidden or evading processes.

\end{document}